\documentclass[fleqn,usenatbib,useAMS]{mnras}

\usepackage{graphicx}	
\usepackage{amsmath}	
\usepackage{amssymb}	
\usepackage{multicol}   
\usepackage{bm}		    
\usepackage{pdflscape}	
\usepackage[dvipsnames]{xcolor}

\usepackage[T1]{fontenc}
\usepackage{ae,aecompl}

\usepackage{newtxtext,newtxmath}

\title[High Frequency Radio Observations of Two Magnetars]{High Frequency Radio Observations of Two Magnetars, PSR J1622$-$4950 and 1E 1547.0$-$5408}

\author[C.-Y.\ Chu et al.]{Che-Yen Chu,$^{1}$\thanks{Contact e-mail: \href{mailto:cychu@gapp.nthu.edu.tw}{cychu@gapp.nthu.edu.tw}}
C.-Y.\ Ng,$^{2}$
Albert K.\ H.\ Kong,$^{1}$
Hsiang-Kuang Chang,$^{1}$
\\
$^{1}$Institute of Astronomy, National Tsing Hua University, Hisnchu, Taiwan \\
$^{2}$Department of Physics, The University of Hong Kong, Hong Kong, China }

\date{Accepted XXX. Received YYY.}

\pubyear{2021}

\begin{document}
\label{firstpage}
\pagerange{\pageref{firstpage}--\pageref{lastpage}}
\maketitle

\begin{abstract}
We investigated the radio spectra of two magnetars, PSR J1622$-$4950 and 1E 1547.0$-$5408, using observations from the Australia Telescope Compact Array and the Atacama Large Millimeter/submillimeter Array taken in 2017. Our observations of PSR J1622$-$4950 show a steep spectrum with a spectral index of $-$1.3 $\pm$ 0.2 in the range of 5.5--45\,GHz during its re-activating X-ray outburst in 2017. By comparing the data taken at different epochs, we found significant enhancement in the radio flux density. The spectrum of 1E 1547.0$-$5408 was inverted in the range of 43--95\,GHz, suggesting a spectral peak at a few hundred gigahertz. Moreover, we obtained the X-ray and radio data of radio magnetars, PSR J1622$-$4950 and SGR J1745$-$2900, from literature and found two interesting properties. First, radio emission is known to be associated with X-ray outburst but has different evolution. We further found that the rising time of the radio emission is much longer than that of the X-ray during the outburst. Second, the radio magnetars may have double peak spectra at a few GHz and a few hundred GHz. This could indicate that the emission mechanism is different in the cm and the sub-mm bands. These two phenomenons could provide a hint to understand the origin of radio emission and its connection with the X-ray properties.
\end{abstract}




\begingroup
\let\clearpage\relax
\endgroup
\newpage

\section{Introduction}
Magnetars are isolated neutron stars with strong magnetic fields $\gtrsim 10^{14}$\,G \citep{Duncan1992}. Unlike rotation-powered pulsars, their X-ray luminosities cannot be solely due to rotational energy loss. Instead, their strong magnetic fields provide the energy to power the observed X-ray emission. Magnetars manifest themselves as soft gamma repeaters (SGRs) or anomalous X-ray pulsars (AXPs) in observations. SGRs and AXPs were first discovered in X-rays as neutron stars with slow rotation periods (1--12\,s) showing burst activities. Currently, there are 24 confirmed magnetars \citep{Olausen2014}\footnote{McGill Online Magnetar Catalog \url{http://www.physics.mcgill.ca/~pulsar/magnetar/main.html}}, and only five of them show pulsed radio emission, of which the radio spectra are either flat or inverted \citep{Camilo2006,Camilo2007a,Levin2010,Eatough2013,Shannon2013,Karuppusamy2020}.

An X-ray outburst of a magnetar is an event of sudden increase in X-ray flux and sometimes comes with multiple X-ray short bursts in millisecond to second timescale. Sudden magnetar crustal activities can twist the magnetosphere resulting in an X-ray outburst as observed \citep{Thompson2008,Beloborodov2009}. The magnetosphere gradually becomes untwisted so that the X-ray flux decreases. The flux of an outburst usually decays in different time scales, with one rapid decay within hours to a day followed by a slow decay in several months to years \citep[e.g.,][]{Woods2004}. Most outbursts are associated with some radiative phenomena such as changes in pulse profile, pulsed fraction and multi-wavelength emission, and X-ray spectral hardening. Multi-wavelength studies of radio-loud magnetars suggest that an X-ray outburst may trigger radio emission \citep[e.g.,][]{Halpern2005,Camilo2007a,Anderson2012}.

The magnetar 1E 1547.0$-$5408 was discovered in a supernova remnant (SNR) G327.24$-$0.13 \citep{Lamb1981} and was confirmed to be an AXP later in X-ray and radio observations with a spin period of 2.1\,s \citep{Gelfand2007,Camilo2007a}. In previous centimeter observations, 1E 1547.0$-$5408 showed a flat spectrum \citep{Camilo2008} which is possible a gigahertz peaked spectrum \citep[GPS;][]{Kijak2013}. The 2007 X-ray observations suggest that there was an X-ray outburst, which could have triggered the radio emission of 1E 1547.0$-$5408 between late 2006 and early 2007 \citep{Halpern2008}. \emph{Swift}/BAT, \emph{Fermi}/GBM and \emph{INTEGRAL}/SPI detected two more outbursts in 2008 October \citep{Israel2010,Kaneko2010} and in 2009 January \citep{Mereghetti2009,Savchenko2010,Kaneko2010}. The 2009 X-ray outburst was more energetic but less variable than the 2008 one \citep{Israel2010,Ng2011,Scholz2011}. There were also some short burst activities in 2009 March and 2010 January \citep{Kaneko2010,Kienlin2012,Kuiper2012}. However, after 2010, there has been no X-ray burst report for 1E 1547.0$-$5408.

PSR J1622$-$4950 is the first magnetar discovered in radio. At the time of discovery, it showed an inverted spectrum from 1.4 to 9.0\,GHz with a spectral index $\alpha = 0.71$ \citep{Levin2010}. The positive index is very rare as radio pulsars generally have indexes ranging from $-$2 to $-$1. However, in later observations, it turned out that the spectrum is not inverted from 17 to 24\,GHz \citep{Keith2011}. PSR J1622$-$4950 shows a GPS similar to the spectra of 1E 1547.0$-$5408 observed in 2007 \citep{Kijak2013}. The X-ray data during the epoch of radio discovery of PSR J1622$-$4950 suggest that an X-ray outburst could have happened before 2007 April \citep{Anderson2012}. Follow up radio observations show that the radio emission decreased to an undetectable level in 2015 \citep{Scholz2017} until the re-activating X-ray outburst event in 2017 March \citep{Pearlman2017,Camilo2018}.

For other magnetars with radio emission detected, XTE J1810$-$197 is the first radio detected magnetar \citep{Camilo2006} that was discovered after its X-ray outburst \citep{Ibrahim2004}. The radio emission disappeared in 2008 and then reactivated during the 2019 X-ray outburst. SGR J1745$-$2900 is a radio magnetar discovered near Sargittarius A* in 2013 \citep{Kennea2013,Mori2013} and it showed the highest frequency radio detection of a magnetar at 291\,GHz \citep{Torne2017}. The newly confirmed magnetar, Swift J1818.0$-$1607, was discovered on 2020 March 12 with a spin period of 1.36\,s \citep{Evans2020,Enoto2020,Esposito2020} and was found to have pulsed radio emisson few days later \citep{Karuppusamy2020,Esposito2020}. The radio emission of Swift J1818.0$-$1607 was detected in a wide band from 0.65 to 154\,GHz. There is another special magnetar, SGR 1935+2154, which was confirmed to be a magnetar in 2014 \citep{Israel2014}. It entered an active state with a forest of short X-ray bursts on 2020 April 27 \citep{Palmer2020}. Later on April 28, a short radio burst was detected from SGR 1935+2154 \citep{CHIME2020,Bochenek2020}. However, pulsed radio emission was detected only at 2 epochs \citep{Burgay2020,Zhu2020}. 

There are two rotation-powered pulsars with high magnetic fields ($\sim 4.5 \times 10^{13}$\,G), PSR J1846$-$0258 and PSR J1119$-$6127 \citep{Gotthelf2000,Camilo2000}, which have shown magnetar-like X-ray outbursts in 2006 and 2016, respectively \citep{Gavriil2008,Archibald2016}. Before and after the outburst, the radio emission of PSR J1119$-$6127 was steady as other radio pulsars. However, during the outburst, the radio emission became variable \citep{Dai2018}. For PSR J1846$-$0258, no radio emission was detected before, during, and after the outburst \citep[e.g.,][]{Gotthelf2000,Archibald2008}. 

Radio magnetars are quite different from radio pulsars. The radio spectra of magnetars are flatter than that of radio pulsars. In contrast to the numerous radio pulsars, only five magnetars have pulsed radio emission. Studying radio magnetars can therefore shed light on the multi-wavelength emission mechanisms of this mysterious compact object. In this paper, we performed high frequency radio observations of two magnetars and compared with their X-ray light curve to investigate their radio spectral behaviours.

\section{Observations and Data Reduction}
\subsection{ATCA}
We observed the two target magnetars with the Australia Telescope Compact Array (ATCA) which consists of six 22-m antennas with the longest baseline of 6\,km. The observations were made in the 3mm and 7mm bands for 1E 1547.0$-$5408 on 2017 August 22 with the array configuration EW352. For PSR J1622$-$4950, we observed in the 7mm, 15mm, 4cm and 16cm bands on 2017 June 24 with the array configuration H214. Mars and PKS 1934$-$638 were observed as flux calibrators for 3mm band and all other bands, respectively. Observation details are described in Table~\ref{tab:observation}.

The data reduction were carried out with MIRIAD \citep{Sault1995}\footnote{See \url{https://www.atnf.csiro.au/computing/software/miriad/}} using standard techniques described in the ATCA Users Guide\footnote{ATCA User Guide \url{https://www.narrabri.atnf.csiro.au/observing/users_guide/html/atug.html}}. After applying bandpass, gain, and flux calibrations, the task \verb'mfclean' was used to clean the entire primary beam and then the flux densities were obtained with the task \verb'imfit'. The noise levels of all frequency band observations are smaller than 0.3\,mJy/beam and the beam sizes are around 4".

\begin{table}
 \caption{Details of ATCA and ALMA observations.}
 \label{tab:observation}
 \begin{tabular}{cccccc}
  \hline
  \hline
  Telescope & Band  & Frequency & Date in   & Observing     & Flux \\
   Array    &       & (GHz)     & 2017      & time (min)    & calibrator \\
  \hline
  ATCA  & 16cm  & 2.1           & Jun 24    & 15    & PKS 1934$-$638 \\
        & 4cm   & 5.5, 9.0      & Jun 24    & 15    & PKS 1934$-$638 \\
        & 15mm  & 16.7, 21.2    & Jun 24    & 40    & PKS 1934$-$638 \\   
        & 7mm   & 43.0, 45.0    & Jun 24    & 50    & PKS 1934$-$638 \\
        &       &               & Aug 22    & 100   & PKS 1934$-$638 \\
        & 3mm   & 93.0, 95.0    & Aug 22    & 120   & Mars \\
  \hline
  ALMA  & 3     & 97.5          & May 19    & 8.6   & J1617$-$5848 \\
        &       &               & Sep 15    & 8.8   & J1617$-$5848 \\
        & 6     & 233           & May 19    & 18.6  & J1617$-$5848 \\
  \hline

 \end{tabular}
\end{table}

\subsection{ALMA}
We observed PSR J1622$-$4950 with the Atacama Large Millimeter/submillimeter Array (ALMA) in Band 3 (97.5\,GHz) and Band 6 (233\,GHz) on 2017 May 19 and September 15. Both bands have usable bandwidth of 7.5\,GHz. We also observed J1617$-$5848 as the bandpass and flux calibrators and J1603$-$4904 as the phase calibrator.

For the May observation, 43 12-m antennas were used with baselines from 15.1\,m to 1.1\,km. The Bands 3 and 6 observations had on-source times of 8.6\,min and 18.6\,min, respectively. For the Sep observation, 39 12-m antennas were used with baselines from 41.4\,m to 9.5\,km. The Band 3 observation had an on-source time of 8.8\,min, but the Band 6 data were not usable due to poor weather condition. Details of the observations are list in Table~\ref{tab:observation}.

We applied the standard data reduction pipeline with CASA v4.7.2 to process the data. We first flagged bad data points, then applied bandpass, gain, and flux calibrations. Finally, total intensity images were constructed and cleaned using the task \verb'tclean'. The Band 3 image taken in May has a beam size of 1.1" and rms noise of 24\,$\mu$Jy/beam. The September one has a beam size of 0.16" and noise of 31\,$\mu$Jy/beam. The Band 6 image from May has a beam size of 0.45" and rms noise of 21\,$\mu$Jy/beam. All the noise levels are consistent with theoretical values.

We did not detect PSR J1622$-$4950 from the May observation, but the source was detected in September at Band 3, with a flux density of 190 $\pm$ 30\,$\mu$Jy.

\begin{table}
 \caption{Results of observations on magnetars.}
 \label{tab:result}
 \begin{tabular}{ccccc}
  \hline
  \hline
  Magnetar  & Date in   & Frequency & Bandwidth & Flux density$^{*}$ \\
            & 2017      & (GHz)     & (GHz)     & (mJy) \\
  \hline
  1E 1547.0$-$5408  & Aug 22    & 43.0  & 1.84  & 6.2 $\pm$ 0.8 \\
                    &           & 45.0  & 1.84  & 6.3 $\pm$ 0.8 \\
                    &           & 93.0  & 1.84  & 8.1 $\pm$ 0.9 \\
                    &           & 95.0  & 1.84  & 9.0 $\pm$ 1.2 \\
  \hline
  PSR J1622$-$4950  & Jun 24    & 2.1   & 1.84  & < 18 \\
                    &           & 5.5   & 1.84  & 10.8 $\pm$ 0.4 \\
                    &           & 9.0   & 1.84  & 3.5 $\pm$ 0.1 \\
                    &           & 16.7  & 1.84  & 3.5 $\pm$ 0.1 \\
                    &           & 21.2  & 1.84  & 1.43 $\pm$ 0.05 \\
                    &           & 43.0  & 1.84  & 0.43 $\pm$ 0.03 \\
                    &           & 45.0  & 1.84  & 0.48 $\pm$ 0.03 \\ 
                    & May 19    & 97.5  & 7.5   & < 0.072 \\
                    &           & 233   & 7.5   & < 0.063 \\
                    & Sep 15    & 97.5  & 7.5   & 0.19 $\pm$ 0.03 \\ 
  \hline
 \multicolumn{5}{l}{\emph{Note.} $^{*}$The upper limit in flux density is 3$\sigma$ detection limit.}
 \end{tabular}
\end{table}

\section{Results}
\subsection{PSR J1622$-$4950}
\label{sec:J1622}

\begin{figure}
 \includegraphics[width=\columnwidth]{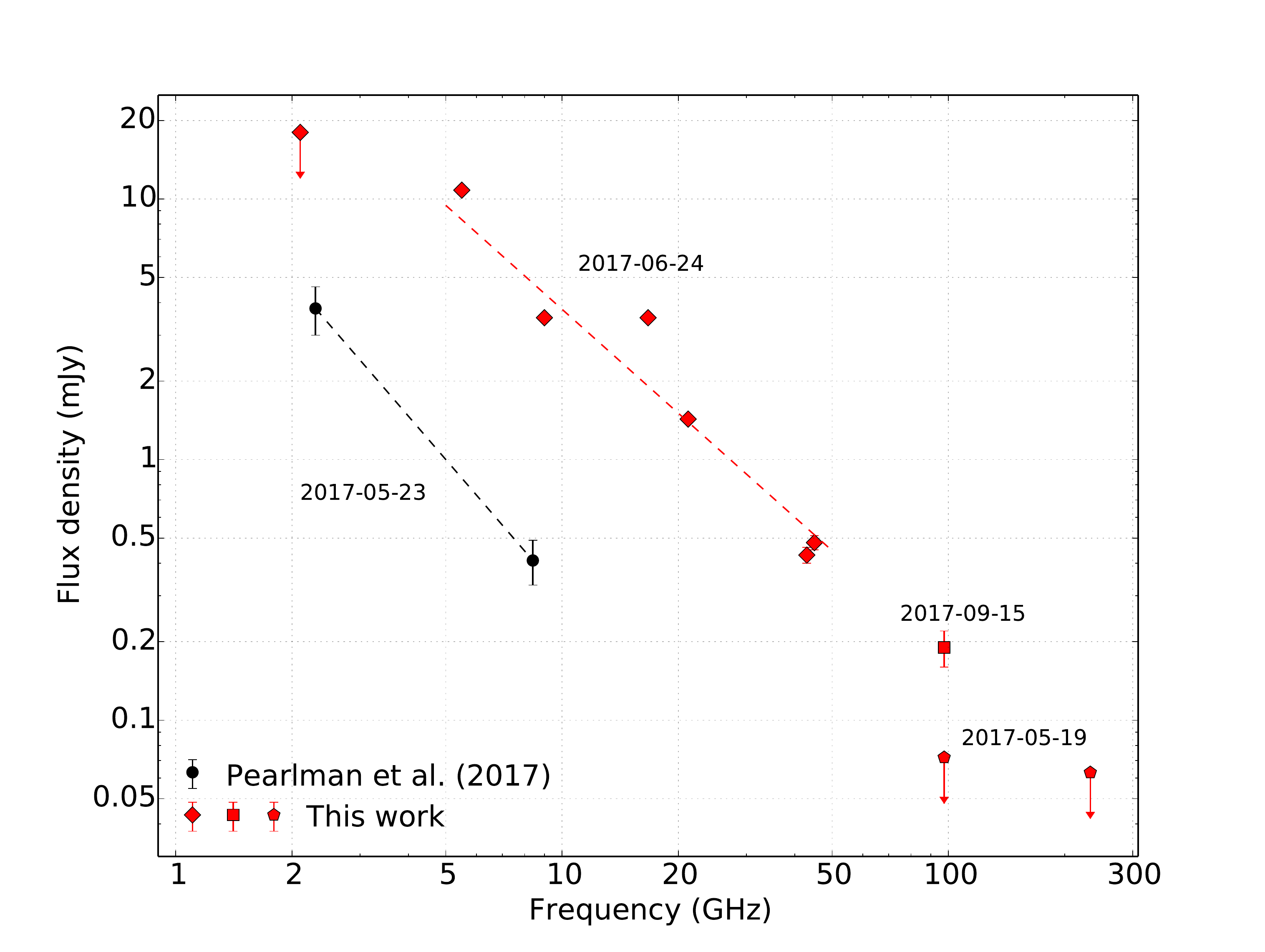}
 \caption{Radio spectra of PSR J1622$-$4950. The black circles are spectrum observed by \citet{Pearlman2017}. The red diamonds are our ATCA measurements and red squares and pentagons are our ALMA measurements. The power-law fitting result in 5.5--45\,GHz is shown in red dashed line and yield a spectral index of $-$1.3 $\pm$ 0.2. For some of the data, the error bars are invisible because the errors are smaller than the size of symbols.}
 \label{fig:spec_j1622}
\end{figure}

The results of ALMA and ATCA observations of PSR J1622$-$4950 are listed in Table~\ref{tab:result} and plotted in Figure~\ref{fig:spec_j1622}. The radio emission was clearly detected from 5.5 to 45\,GHz by ATCA. We therefore fit the steep radio spectrum between 5.5 and 45\,GHz with a power-law and yield a spectral index of $-$1.3 $\pm$ 0.2. Our ALMA observations at 97.5 and 233\,GHz showed non-detection with a 3$\sigma$ upper limit of 0.08\,mJy on May 19. However, the flux density at 97.5\,GHz increased to 0.19\,mJy after 4 months.

The black points in Figure~\ref{fig:spec_j1622} are taken from \citet{Pearlman2017}. Their observations on May 23 in 2017 yield a flux density of 3.8/0.41\,mJy at 2.3/8.4\,GHz, respectively. Our observation taken one month later, June 24, however shows a flux density of 10.8 and 3.5\,mJy at 5.5 and 9.0\,GHz, respectively. The cm band flux density increased by nearly an order of magnitude in only one month. Our ALMA observations in Band 3 also show an increase in flux density (see Section~\ref{sec:flux} for discussion).

\subsection{1E 1547.0$-$5408}

\begin{figure}
 \includegraphics[width=\columnwidth]{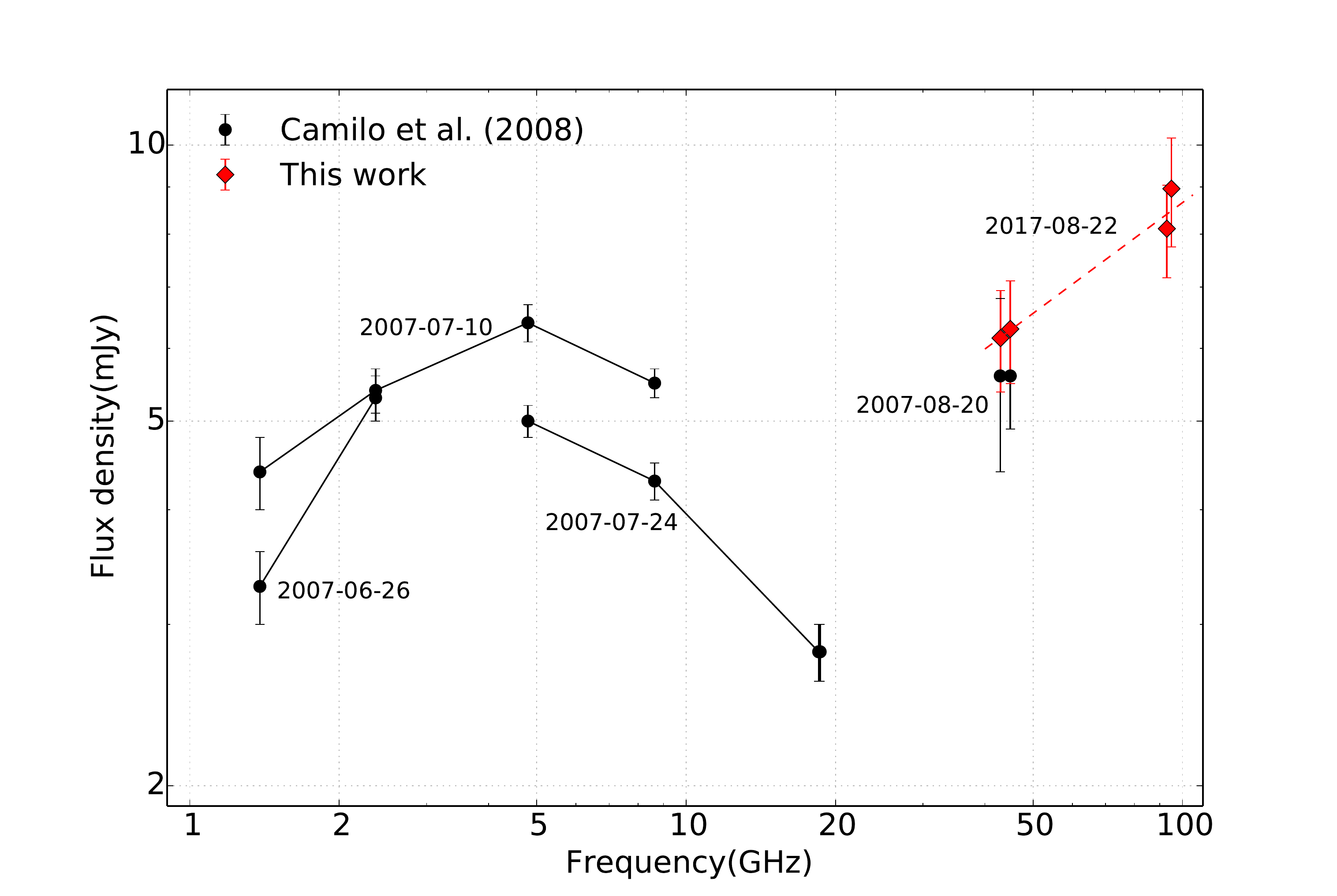}
 \caption{Radio spectra of 1E 1547.0$-$5408. The black circles connected by lines show the data in different day taken from \citet{Camilo2008}. The red diamonds are our ATCA observation results fitted with power-law shown in dashed line. Our results show an inverted spectrum from 43 to 95\,GHz with a spectral index of 0.4 $\pm$ 0.1.}
 \label{fig:spec_1e1547}
\end{figure}

Our ATCA measurements of 1E 1547.0$-$5408 yield flux densities of 6.2\,mJy at 43\,GHz, 6.3\,mJy at 45\,GHz, 8.1\,mJy at 93\,GHz and 9.0\,mJy at 95\,GHz. The results are listed in Table~\ref{tab:result} and plotted in Figure~\ref{fig:spec_1e1547}. The spectrum is fit with a power-law and we found a positive spectral index of 0.4 $\pm$ 0.1. The inverted spectrum from 43 to 95\,GHz is different from the flat spectrum in the cm band, indicating a possible spectral peak at high frequency (see Section~\ref{sec:2peak} for discussion).

The long-term X-ray light curve of 1E 1547.0$-$5408 shows that the absorbed X-ray flux has gradually decreased since the 2009 outburst but the flux level in 2017 remained much higher than the lowest flux level in 2006 \citep[Fig. 3 in][]{CotiZelati2020}. The X-ray flux during our ATCA observations in 2017 was higher than the X-ray flux in mid-2007, the epoch of radio observations taken by \citet{Camilo2008}.

\section{Discussion}
\subsection{Flux evolution}
\label{sec:flux}

\begin{figure*}
 \includegraphics[width=\textwidth]{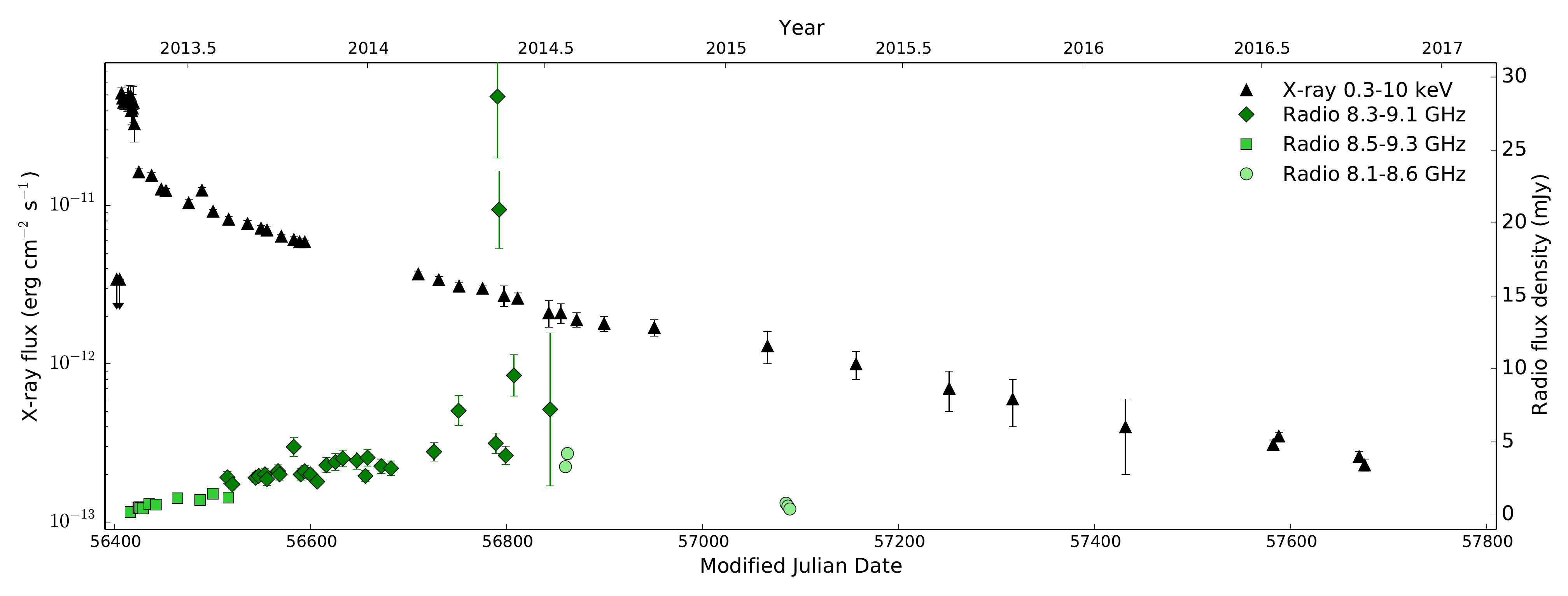}
 \caption{X-ray and radio light curves of SGR J1745$-$2900. The black triangles are absorb 0.3--10\,keV X-ray flux taken from \citet{Kennea2013} and \citet{CotiZelati2015,CotiZelati2017}. The green diamonds, squares and circles are X band radio flux density taken from \citet{Lynch2015}, \citet{Pennucci2015}, and \citet{Torne2015,Torne2017}. For some of the data, the error bars are invisible because the errors are smaller than the size of symbols.}
 \label{fig:lc_j1745}
\end{figure*}

The radio emission of magnetars is known to be variable \citep[e.g.,][]{Camilo2006,Levin2010} and it has different evolution than the X-ray emission \citep{Lynch2015,Pennucci2015}. We obtained radio and X-ray data from literature and found that the radio emission has longer rising time than X-ray emission. The most obvious evidence is the 2013 outburst of SGR J1745$-$2900.

The X-ray and the X band (8.1--9.3\,GHz) radio light curves of SGR J1745$-$2900 taken from literature are shown in Figure~\ref{fig:lc_j1745}. The radio emission started to decrease about $\sim$ 400 days after the outburst, giving a radio rising time of 400 days. In contrast, the X-ray outburst is an event of sudden increase in the persistent X-ray flux and it decayed rapidly after the first day. The rising time of radio emission from SGR J1745$-$2900 is then much longer than that of X-ray emission.

\begin{figure*}
 \includegraphics[width=\textwidth]{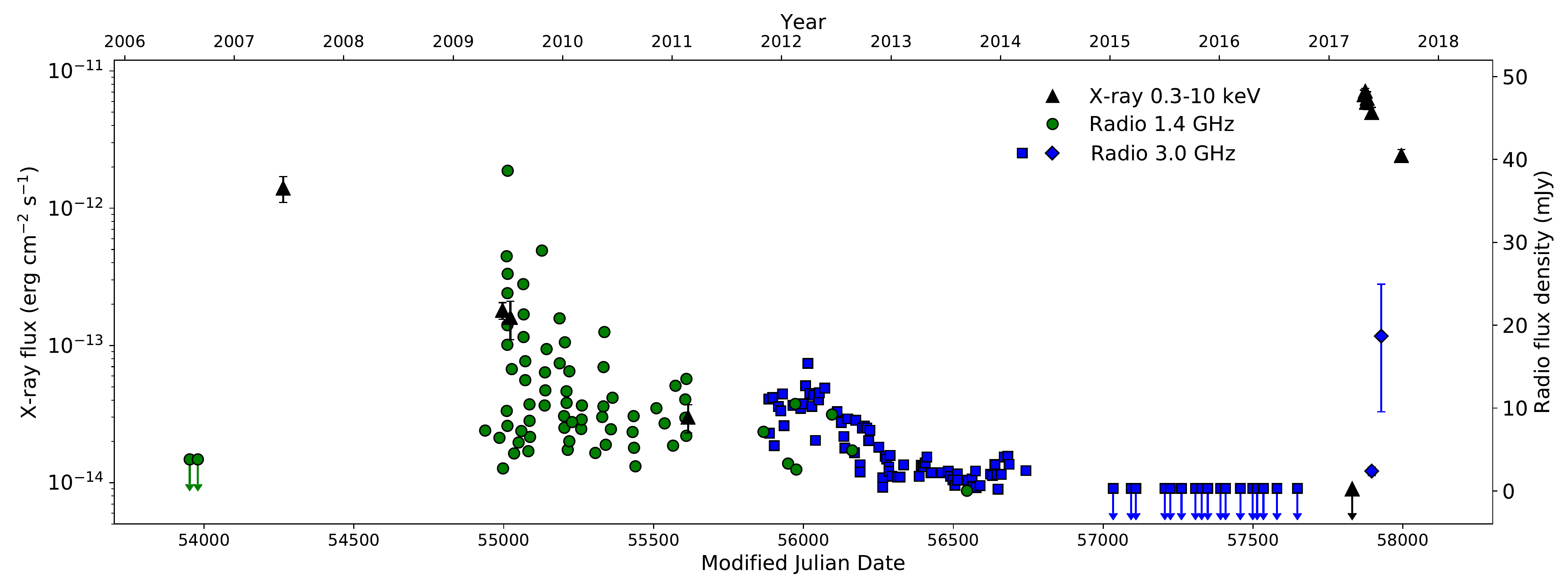}
 \caption{X-ray and radio light curves of PSR J1622$-$4950. The black triangles are absorb 0.3--10\,keV X-ray flux taken from \citet{Anderson2012} and \citet{Camilo2018}. The green circles and the blue squares are respectively 1.4\,GHz and 3\,GHz radio flux density taken from \citet{Scholz2017}. The blue diamonds are 3\,GHz radio flux density derived from \citet{Pearlman2017} and this work (Figure~\ref{fig:spec_j1622}) based on the observed spectral indices. For some of the data, the error bars are invisible because the errors are smaller than the size of symbols.}
 \label{fig:lc_j1622}
\end{figure*}

As described in Section~\ref{sec:J1622}, the radio flux density of PSR J1622$-$4950 showed a very significant increase while the X-ray flux was decaying in the 2017 outburst, i.e., the rising time is different in radio and X-rays. Radio emission had longer rising time than X-ray emission in 2017 outburst of PSR J1622$-$4950. Figure~\ref{fig:lc_j1622} shows the 0.3--10\,keV X-ray light curve and 1.4 and 3.0\,GHz radio light curves of PSR J1622$-$4950. According to the X-ray and radio observations, there should be an outburst event between 2006 September and 2007 April \citep{Anderson2012}. The 1.4 and 3.0\,GHz observations during this outburst showed very strong variability. Although the trend of radio emission was monotonic decrease and then went down to undetectable level in 2015, since there is no data from 2007 to 2009, we can not rule out that the radio emission is rising during this time.

Moreover, the newly discovered magnetar, Swift J1818.0$-$1607, showed significant rising trend in radio emission during its 2020 March outburst. The S band (2.0--2.3\,GHz) and the X band (8.3--8.8\,GHz) radio flux densities increased from 0.05 to 0.7\,mJy in three epochs from March to July \citep{Maan2020,Majid2020a,Majid2020b} and from 0.026 to 1.29\,mJy in four epochs from March to August \citep{Majid2020a,Majid2020b,Pearlman2020,Ding2020}, respectively. The increasing trend of radio flux density suggests that the radio rising time is $\gtrsim$ 150 days, which is much longer than the X-ray rising time in outburst \citep{Hu2020}.

For 1E 1547.0$-$5408, there are only a few radio flux density measurements so we only discuss the 2007 observations done by \citet{Camilo2008}. From Figure~\ref{fig:spec_1e1547}, we can see that the flux density increased from June 26 to July 10 at 1.4\,GHz and then showed a decreasing trend at 4.8--8.6\,GHz two weeks later in July 24. The spectral evolution here gives us another example that the cm band spectrum seems to have a longer rising time than in X-rays during an outburst. However, the increase of flux density can also be explained by the variability since the radio emission of magnetars is highly variable.

Furthermore, during the 2003 outburst of XTE J1810$-$197, the X-ray flux evolved as a typical magnetar X-ray outburst, whit a rapid increase in flux followed by different scale of decay \citep{Alford2016}. The radio flux densities measured in 2003, 2004, and early 2006 were all lower than the measurements in mid-2006 \citep{Halpern2005,Camilo2006,Camilo2016}, indicating that there could be a very long radio rising period between 2004 and 2006. However, during the 2018 X-ray outburst \citep{Gotthelf2019}, the radio light curve showed no significant increase \citep{Levin2019}. The rising time of the radio emission was then fewer than 22 days since outburst, which is much shorter than in other cases.

\citet{Rea2012} pointed out that the emergence of radio emission from magnetars has a delay after the X-ray outburst. This could be attributed to the twisted magnetosphere. The twist and the magnetospheric charge are so strong when X-ray flux reaching the peak \citep{Thompson2002} that the particle cascades, which emit radio emission, could not easily propagate outward. But the 2013 outburst of SGR J1745$-$2900 showed that radio emission can be seen at the very beginning \citep{Pennucci2015}. The delay of radio emission can then be better described as slow rising in radio emission. During the X-ray flux decay, the magnetosphere gradually untwists so that particles has a higher chance to propagate outside the magnetosphere. Hence, the radio emission gradually increases, showing a longer rising time scale than the X-ray flux. 

In addition to radio magnetars, the radio pulsar with magnetar-like bursts, PSR J1119$-$6127, also showed difference in X-ray and radio flux evolution. PSR J1119$-$6127 had an X-ray outburst on July 26 in 2016. The radio flux density of PSR J1119$-$6127 was steadily to be $\sim$ 1\,mJy at 1.4\,GHz before this 2016 outburst. However, the radio emission became variable after the outburst. Observations showed that the radio emission disappeared after July 29 \citep{Burgay2016a}. On August 9, the radio emission re-activated but the flux density was significantly weaker than the pre-outburst value \citep{Burgay2016b}. After the re-activation, the radio flux density gradually increased to its highest value of $\sim$ 5.6\,mJy on August 31, one month after the X-ray outburst \citep[Fig. 1 in][]{Dai2018}. The radio flux density then dropped to the minimum value of $\sim$ 0.14\,mJy and then recovered to PSR J1119$-$6127's steady flux density of 1\,mJy. The variable radio emission spent one month to reach its highest value during an X-ray outburst. This longer rising time is similar to other radio magnetars. The variable state of radio emission causing by X-ray outburst can then be described by a magnetar model. In normal rotation-powered radio pulsars, the radio emission is believed to be generated by particles in open magnetic field lines near the polar region. In the case of PSR J1119$-$6127, before and after the X-ray outburst, it reacted as rotation-powered pulsar. The radio emission comes from radio beam steadily. At the time of X-ray outburst, the neutron star's activity make the magnetosphere twisted so that the original open field lines become unstable \citep{Beloborodov2009,Braithwaite2004}. Hence, the radio emission from radio beam disappeared. Similarly, the magnetosphere of PSR J1622$-$6127 changed to a magnetar-like structure. Radio emission gradually increased due to magnetosphere untwisting. At the end of the outburst, the magnetosphere gradually re-configurated back to the structure of normal rotation-powered radio pulsars so that the radio emission became stable again. The difference of radio pulse profile during radio variable state and steady state \citep{Archibald2017,Majid2017,Dai2018} can also support the transition between a usual radio pulsar model and a radio magnetar model.

All the five radio magnetars and one magnetar-like rotation-powered pulsar show hints of rising time difference between X-ray and radio emission. Although not all the outburst show similar phenomenon, the evolution difference can provide more insight to study the relation between X-ray and radio emission from magnetars during an outburst.

\subsection{Double peak spectra}
\label{sec:2peak}

The observed radio spectra of 1E 1547.0$-$5408 and PSR J1622$-$4950 show a peak at a few GHz \citep{Camilo2008,Levin2010,Keith2011}. This spectral feature is so-called gigahertz-peaked spectrum (GPS) which is used to describe the spectral turn-over at a few GHz for some pulsars \citep{Kijak2011}. Detailed spectral analysis shows that both spectra can be fit with a log-parabola and the peak is around 5.0\,GHz for 1E 1547.0$-$5408 and 8.3\,GHz for PSR J1622$-$4950 \citep{Kijak2013}. The fourth discovered radio magnetar, SGR J1745$-$2900, also shows GPS with a peak at $\sim$ 2\,GHz \citep{Pennucci2015}. Moreover, simultaneous wide band observation from 0.7 to 4\,GHz for XTE J1810$-$197 shows an inverted spectrum \citep{Dai2019}. Compared to its past higher frequency observations, XTE J1810$-$197 is likely to have GPS feature as well with a peak at $\sim$ 4\,GHz. The reason for pulsars having GPS is likely the low frequency absorption from the surrounding environment of pulsars or the molecular clouds in the line of sight \citep{Kijak2011}. Both 1E 1547.0$-$5408 and PSR J1622$-$4950 are in SNRs \citep{Lamb1981,Gelfand2007,Levin2010,Anderson2012} and SGR J1745$-$2900 is only 2" away from the Galactic center \citep{Kennea2013}. The X-ray, optical and infrared (IR) image of XTE J1810$-$497 shows no evidence of association with other source \citep{Gotthelf2004}. If the 0.7--4\,GHz band spectrum of XTE J1810$-$197 is actually a GPS, the absorption could happen in the molecular clouds in the line of sight or the reason for the GPS from magnetars may be different from that of the radio pulsars.

Moreover, the spectra of radio magnetars could have not only a single peak at GHz but also a second peak at few hundred GHz. Observations of SGR J1745$-$2900 in 2015 show an inverted spectrum from 2.54 to 291\,GHz \citep{Torne2017}. However, not all the data points are well fit. The cm band (2.54--8.35\,GHz) spectral index is negative and the flux density likely decreases above 200\,GHz. The spectral shape can then be better described by multiple components. Considering the GPS at lower frequency, the observed spectral shape in 2015 may have double peaks at $\sim$ 2 and $\sim$ 200\,GHz. Comparing spectra of SGR J1745$-$2900 at different epochs from 2014 to 2015, the cm band flux density decreased by about 5 times while sub-mm band flux density increased about 4 times \citep{Torne2015,Torne2017}. This evolution difference supports the theory of a double-component radio spectrum.

In addition, the observed inverted spectrum of 1E 1547.0$-$5408 between 43 and 95\,GHz gives us a hint of a peak over 100\,GHz (Figure~\ref{fig:spec_1e1547}). By combining observations in the cm and the mm bands, 1E 1547.0$-$5408 shows similar double peak spectrum with SGR J1745$-$2900. The nature of hundred gigahertz peaked spectra remains unknown. The double peaks in the spectrum are likely the result of two different emission mechanisms at different bands. Radio emission from pulsars is known to be coherent, which is characterized by their high brightness temperature. On the other hand, observations at IR and higher energies show that the pulsar emission is incoherent \citep[e.g.,][]{Shklovsky1970}, implying a transition between coherence and incoherence. It has been speculated that the transition should lie in the range of millimeter/submillimeter \citep{Michel1982}. The spectra of both 1E 1547.0$-$5408 and SGR J1745$-$2900 show that the transition between two peaks locate at $\sim$ 20--40\,GHz where could be the end of coherent emission and the start of incoherent emission. There is only one example of double peak spectra from SGR J1745$-$2900 and one candidate from 1E 1547.0$-$5408. More high frequency radio observations are needed to discuss and verify the double peak spectra.

However, our ALMA observations of PSR J1622$-$4950 do not indicate any peak feature at a few hundred GHz. The high frequency 88.5\,GHz observations of XTE J1810$-$197 in 2006 and 2007 also show no hint of hundred-gigahertz-peaked feature \citep{Camilo2007b}. Since the hundred-gigahertz peak of SGR J1745$-$2900 in 2014 was $\sim$ 4 times weaker than in 2015 \citep{Torne2015,Torne2017}, there is still possibility that the high frequency component of PSR J1622$-$4950 and XTE J1810$-$197 was not dominate at the time of their observations.

\subsection{Turning-off of radio emission}

Previously observed radio emission from magnetars can all be associated with an X-ray outburst except for the archival radio data of PSR J1622$-$4950 before 2003 obtained by \citet{Levin2010} because of the lack of X-ray data. The radio emission from XTE J1810$-$197 and PSR J1622$-$4950 disappeared years after an X-ray outburst \citep{Camilo2016,Scholz2017}. Although the last X-ray outburst of 1E 1547.0$-$5408 was in 2009, 8.5 years before our radio observation, we can still have detection in the mm band. There are three possible explanations. The first is that once the radio emission was turned on, it needs to be turned off by other activities. We consider this less likely since there was no evidence of unusual change in some properties such as spin frequency or X-ray pulse profile when the radio emission was undetectable \citep{Camilo2016,Scholz2017}. Another possibility is simply that the 2009 X-ray outburst have not ended so that the X-ray-outburst-associated radio emission continued. The last one is that the radio emission is not associated with X-ray outburst but the X-ray flux level. The X-ray flux level in 2017 is higher than that in 2007, when the radio emission of 1E 1547.0$-$5408 was discovered \citep{CotiZelati2020}, so we can still detect radio emission. Once the X-ray flux decreases to certain level, the radio emission may disappear. The study of \citet{Camilo2016} of XTE J1810$-$197 also suggests that the radio disappearance is most likely due to the decrease of X-ray flux. But they emphasize that the X-ray flux change is only 20\%. More observations on X-ray and radio and more timing and/or spectral analysis are needed to confirm the radio turn-off criteria.

\section{Summary}
We performed analysis of radio observations of two magnetars, 1E 1547.0$-$5408 and PSR J1622$-$4950. The former showed an inverted spectrum from 43 to 95\,GHz, indicating a peak at high frequency. The broadband radio spectra of 1E 1547.0$-$5408 is similar to that of SGR J1745-2900. Both of them may have different emission mechanisms at cm and sub-mm band resulting double peak spectra with peak at a few GHz and a few hundred GHz. Our observations of PSR J1622$-$4950 found a steep spectrum from 5.5 to 45\,GHz. By comparing observations in different epochs, PSR J1622$-$4950 showed a significant increase in radio flux density while the X-ray flux was decreasing. We further obtained the X-ray and radio data of radio magnetars and a magnetar-like radio pulsar from literature and found, for the first time, that the rising time of radio emission is much longer than that of X-ray emission in some cases of magnetar outburst. More simultaneous broadband observations in different epochs are needed to confirm these properties and to study the origin of radio emission from magnetars.  

\section*{Acknowledgements}
\addcontentsline{toc}{section}{Acknowledgements}
The Australia Telescope Compact Array is part of the Australia Telescope National Facility which is funded by the Australian Government for operation as a National Facility managed by CSIRO. This paper makes use of the following ALMA data: ADS/JAO.ALMA\#2016.1.00456.T. ALMA is a partnership of ESO (representing its member states), NSF (USA) and NINS (Japan), together with NRC (Canada), MOST and ASIAA (Taiwan), and KASI (Republic of Korea), in cooperation with the Republic of Chile. The Joint ALMA Observatory is operated by ESO, AUI/NRAO and NAOJ. C.-Y.\ Chu and H.-K.\ Chang are supported by the Ministry of Science and Technology (MOST) of the Republic of China (Taiwan) under the grant MOST 108-2112-M-007-003. C.-Y.\ Ng is supported by a GRF grant from the Hong Kong Government under HKU 17300215P. A.\ K.\ H.\ Kong is supported by the MOST of ROC (Taiwan) under the grants MOST 105-2119-M-007-028-MY3 and 108-2628-M-007-005-RSP. 

\section*{Data availability}
\addcontentsline{toc}{section}{Data availability}
The data underlying this article are available in the article. ATCA data are available through Australia Telescope Online Archive\footnote{\url{https://atoa.atnf.csiro.au/}}. ALMA data are available through ALMA Science Archive\footnote{\url{https://almascience.eso.org/asax/}}.

\bibliographystyle{mnras}
\bibliography{magnetar}


\bsp	
\label{lastpage}
\end{document}